\documentclass[a4paper,aps,prl,showpacs,twocolumn,floatfix]{revtex4-1}

\usepackage{amsmath,amssymb,amsbsy}
\usepackage{graphicx}
\usepackage{hyperref}
\usepackage{xspace}
\usepackage{float}
\usepackage{verbatim}
\usepackage{rotating}
\usepackage{multirow}

\newcommand{\MeV}{\,\ensuremath{\text{MeV}}\xspace}
\newcommand{\cosec}{\operatorname{cosec}}

\begin{document}

\title{Novel insights into transfer processes in the reaction $^{16}$O+$^{208}$Pb at sub-barrier energies}
\author{M. Evers}
\email{maurits.evers@anu.edu.au}
\author{C. Simenel}
\author{M. Dasgupta}
\author{D. J. Hinde}
\author{D. H. Luong}
\author{R. Rafiei}
\author{R. du Rietz}
\affiliation{Department of Nuclear Physics, Research School of Physics and Engineering, Australian National University, Canberra, Australian Capital Territory 0200, Australia}

\date{\today}

\begin{abstract}
The collision of the doubly-magic nuclei $^{16}$O+$^{208}$Pb is a benchmark in nuclear reaction studies. Our new measurements of back-scattered projectile-like fragments at sub-barrier energies show show that transfer of 2 protons ($2p$) is much more probable than $\alpha$-particle transfer. $2p$ transfer probabilities are strongly enhanced compared to expectations for the sequential transfer of two uncorrelated protons; at energies around the fusion barrier absolute probabilities for two proton transfer are similar to those for one proton transfer. This strong enhancement indicates strong $2p$ pairing correlations in $^{16}$O, and suggests evidence for the occurrence of a nuclear supercurrent of two-proton Cooper pairs in this reaction, already at energies well below the fusion barrier.
\end{abstract}


\maketitle

Collisions of heavy ions at energies well below and close to the fusion barrier are entirely driven by quantum mechanics. For example, sub-barrier fusion occurs through quantum tunnelling of the projectile nucleus through the fusion barrier. This process in turn is affected by the internal structure of the collision partners \cite{Dasgupta98}, leading to a coherent superposition of reaction channels. Amongst the possible reactions competing with fusion, transfer of more than one nucleon is certainly the least well understood mechanism, and constitutes an important task to be described both experimentally and theoretically. In particular, the distinction between sequential and cluster transfer is a great challenge, not only in nuclear physics \cite{Oertzen01}, but also in electron transfer between ions or atomic cluster collisions \cite{Campbell00}. In nuclear collisions, the transfer of a cluster of nucleons is a clear signature of correlations between the transferred nucleons affecting the dynamics. Pairing between nucleons of the same isospin as well as $\alpha$-particle clustering have been considered as the most important correlations affecting multi-nucleon transfer \cite{Oertzen01}.\par

Measurements of transfer probabilities in various reactions and at energies near the fusion barrier have therefore been utilized to investigate the role of pairing correlations between the transferred nucleons. Pairing effects are believed to lead to a significant enhancement of pair and multi-pair transfer probabilities \cite{Broglia78,Chiodi82,Mermaz96,Oertzen01,Corradi09}. Closely related to the phenomenon of pairing correlations is the nuclear Josephson effect \cite{{Goldanskii65,*Dietrich71}}, which is understood as the tunneling of nucleon pairs (i.e. nuclear Cooper-pairs) through a time-dependent barrier at energies near but below the fusion barrier. This effect is believed to be similar to that of a supercurrent between two superconductors separated by an insulator. An enhancement of the transfer probability at sub-barrier energies is therefore commonly related to the tunneling of (multi-)Cooper-pairs from one superfluid nucleus to the other \cite{Oertzen01}.\par

The reaction $^{16}$O+$^{208}$Pb can be considered a benchmark in low-energy heavy-ion collisions \cite{deVries75,Hasan79,Thompson89,Tamura80,Becchetti74,Goetz75,Videbaek77,Simenel10}. In the independent particle shell model, both nuclei are doubly-magic with a closed shell of protons and neutrons. However, one and two proton knockout measurements of $^{16}$O using inelastic electron scattering indicate that pairing correlations in the $^{16}$O nucleus lead to a reduction of the spectroscopic factors of excited states just below the Fermi surface \cite{Onderwater97,*Hesselink00,*Dickhoff10}. Furthermore, recent results based on the time-dependent density matrix approach \cite{Assie09} indicate that pairing correlations may indeed lead to cluster structure effects in $^{16}$O. Various sources \cite{Abdullah03,Dorsch08} report on excited states in $^{16}$O and other oxygen isotopes, strongly supporting an $\alpha$-cluster structure.\par

Extensive measurements probing different transfer channels for the $^{16}$O+$^{208}$Pb reaction exist. It was commonly believed that the dominant transfer process involving the exchange of two charged nucleons at energies near the fusion barrier was $\alpha$-particle transfer ($^{16}$O,$^{12}$C) \cite{{deVries75,*Hasan79,*Thompson89,*Tamura80}}. While at energies well above the fusion barrier, $2p$ transfer ($^{16}$O,$^{14}$C) in the same reaction was also observed \cite{Becchetti74,Goetz75,Videbaek77}, relative probabilities between $\alpha$-particle and $2p$ transfer were not addressed.\par

This letter presents evidence that (1) $2p$ transfer (and not $\alpha$-particle transfer) is the dominant transfer process leading to $\Delta Z=2$ events in the reaction $^{16}$O+$^{208}$Pb at energies well below the fusion barrier, and (2) $2p$ transfer is significantly enhanced compared to predictions assuming the sequential transfer of uncorrelated protons, with absolute probabilities as high as those of $1p$ transfer at energies near the fusion barrier.

\begin{figure}[!tb]
\centering
\includegraphics[scale=0.44]{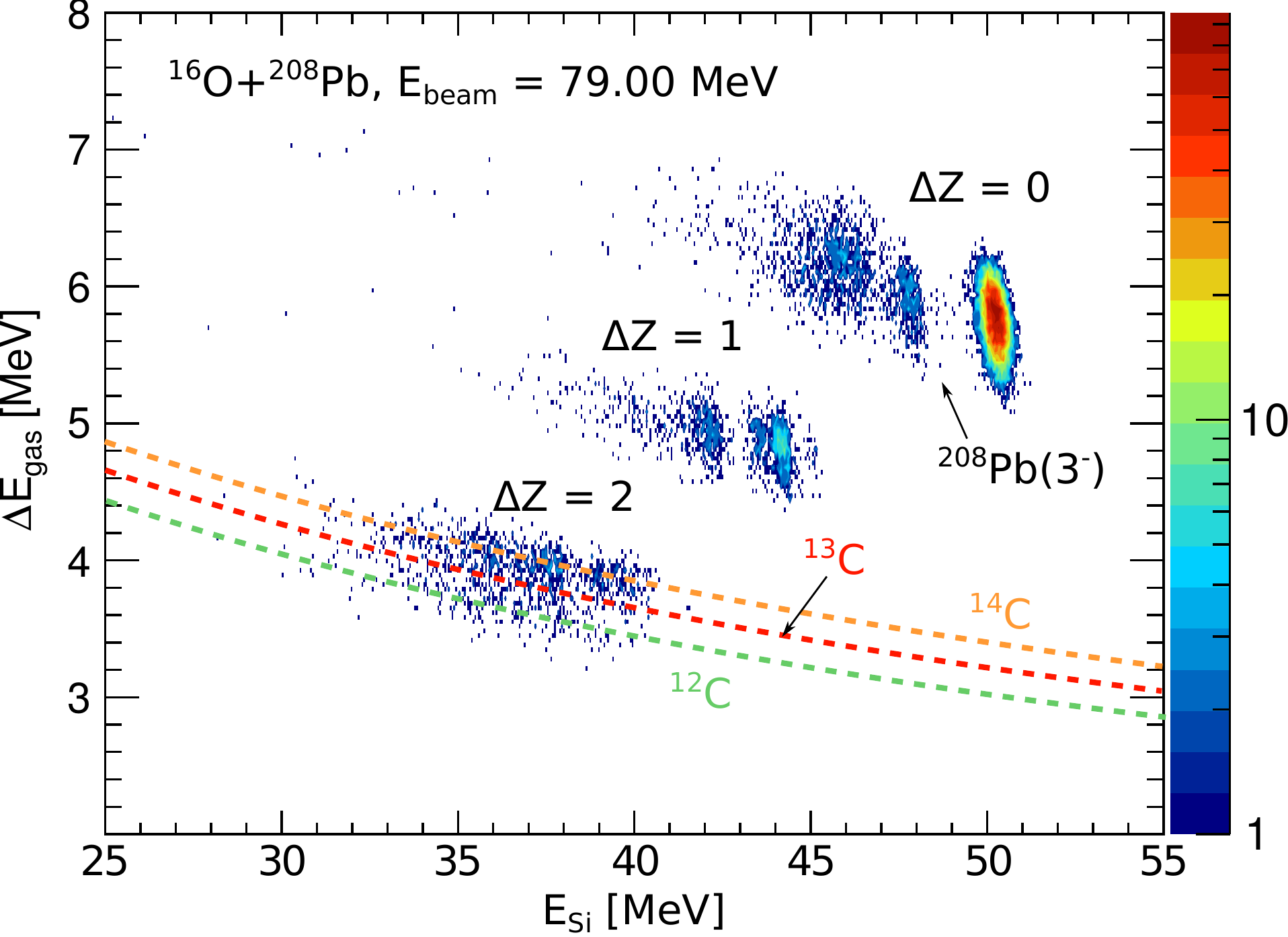}
\caption{(Color online) Two dimensional $\Delta E-E$ spectrum for the reaction $^{16}$O+$^{208}$Pb at the indicated beam energy, corresponding to $E_{c.m.}/V_B = 0.98$. The three different regions indicating $\Delta Z=0,1,2$ transfer are labeled. Calculated energy loss curves (dashed curves) for $^{14}$C, $^{13}$C and $^{12}$C are shown (see text).\label{fig:comparison}}
\end{figure}
Measurements were carried out using the 14UD electrostatic accelerator of the Australian National University. Beams of $^{16}$O were incident on a $^{208}$PbS target with a thickness of 100 $\mu$g/cm$^2$, evaporated onto a 15 $\mu$g/cm$^{2}$ C backing. A detector telescope consisting of a gas ionization chamber and a Si detector located at a backward angle of $\theta_\text{lab}=162^\circ$ was used to record the energy and energy loss of the back-scattered projectile-like fragments (PLFs). Two Si monitors positioned at $\pm 30^\circ$ were used to normalize the back-scattered events to the Rutherford cross section. A typical two dimensional spectrum at a beam energy corresponding to $E_{c.m.}/V_B = 0.98$ is shown in Fig.~\ref{fig:comparison}. The three distinct regions correspond to oxygen, nitrogen and carbon PLFs, which are associated with the transfer of $\Delta Z=0$, $1$ and $2$ units of charge. The main peak at $E_{\text{Si}}\sim 50\MeV$ corresponds to elastically scattered $^{16}$O particles. Events resulting from the transfer of three or more charged nucleons ($\Delta Z\ge 3$) are not observed for measurements at sub-barrier energy.\par

\begin{figure}[!tb]
\centering
\includegraphics[scale=0.55]{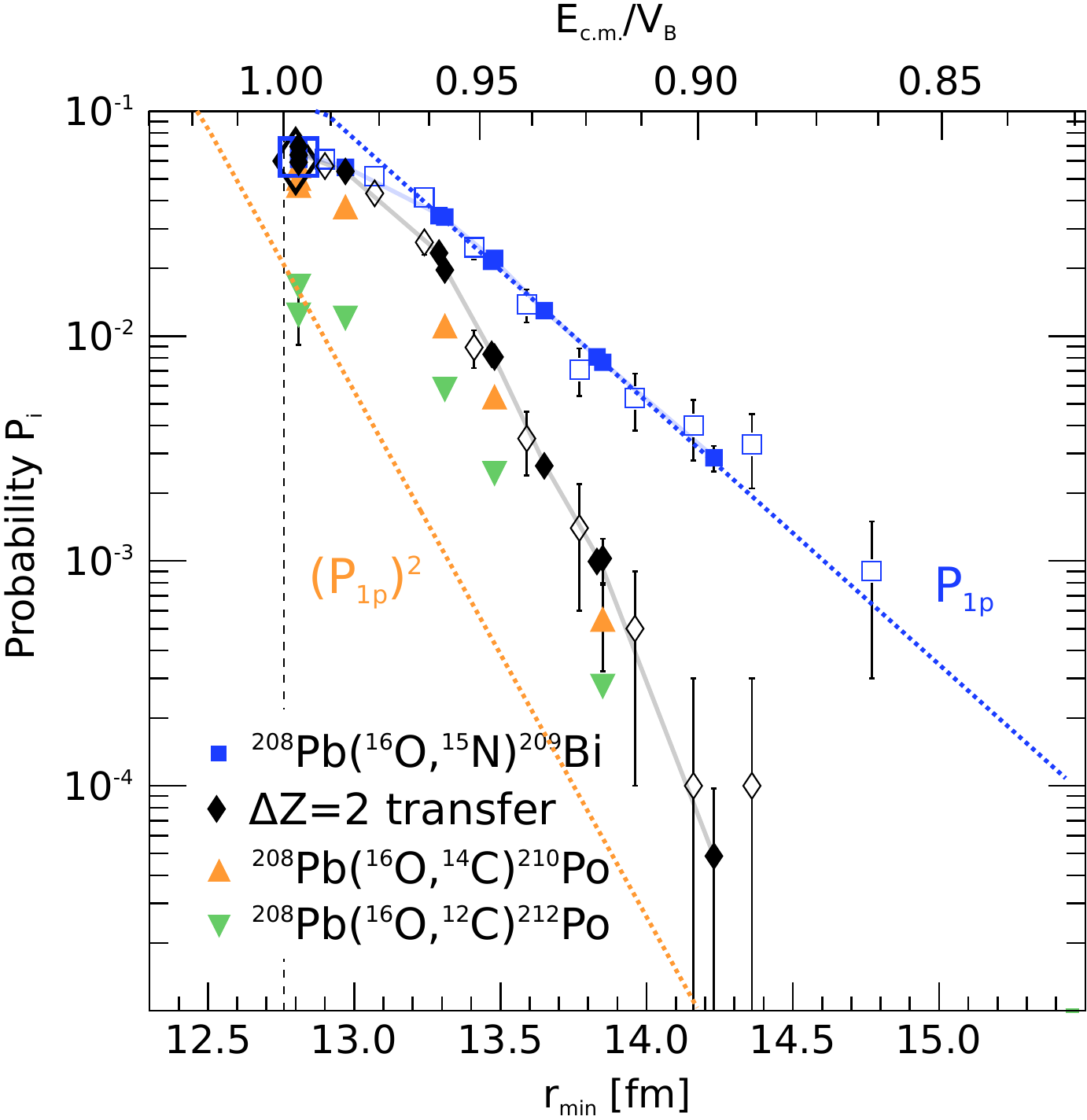}
\caption{(Color online) Transfer probabilities for the indicated transfer processes as a function of the distance of closest approach, see Eq.~\eqref{eq:rc}. The asymptotic behaviour for $1p$ transfer and sequential $2p$ transfer are shown by the dotted straight lines. The vertical dashed line indicates the barrier position. The large open square and diamond at $E_{c.m.}/V_B\sim 1.0$ are the measurements for N (blue) and C PLFs (black) from Videbaek et al. \cite{Videbaek77}. The smaller open squares and diamonds are the measurements for N (blue) and C PLFs (black) from Timmers \cite{Timmers96}.\label{fig:transfer_rmin}}
\end{figure}
\begin{table}[!tb]
\caption{Reaction ground state $Q$-values for selected transfer processes in the reaction $^{16}$O+$^{208}$Pb. Processes with a plus sign correspond to pickup, a minus sign indicates stripping. Predominant processes as determined by our measurements and previous work \cite{Videbaek77,Pieper78} are highlighted in bold.\label{tab:qvalues}}
\begin{tabular}{p{0.5cm}lcc}\hline\hline
 & Reaction & Process & $Q_\text{gs}$ [MeV]\\\hline
\multirow{3}{*}{\begin{sideways}$\Delta Z=0$\end{sideways}} & $^{208}$Pb$(^{16}$O,$^{17}$O$)^{207}$Pb & $\boldsymbol{+1n}$ & -3.225 \\
 & $^{208}$Pb$(^{16}$O,$^{18}$O$)^{206}$Pb & $\boldsymbol{+2n}$ & -1.918 \\
 & $^{208}$Pb$(^{16}$O,$^{15}$O$)^{209}$Pb & $-1n$ & -11.727 \\\hline
\multirow{3}{*}{\begin{sideways}$\Delta Z=1$\end{sideways}} & $^{208}$Pb$(^{16}$O,$^{15}$N$)^{209}$Bi & $\boldsymbol{-1p}$ & -8.328 \\
 & $^{208}$Pb$(^{16}$O,$^{14}$N$)^{210}$Bi & $-1p-1n$ & -14.557 \\
 & $^{208}$Pb$(^{16}$O,$^{16}$N$)^{208}$Bi & $-1p+1n$ & -13.299 \\\hline
\multirow{5}{*}{\begin{sideways}$\Delta Z=2$\end{sideways}} & $^{208}$Pb$(^{16}$O,$^{14}$C$)^{210}$Po & $\boldsymbol{-2p}$ & -13.553 \\
 & $^{208}$Pb$(^{16}$O,$^{13}$C$)^{211}$Po & $-2p-1n$ & -17.178 \\
 & $^{208}$Pb$(^{16}$O,$^{12}$C$)^{212}$Po & $\boldsymbol{-2p-2n}$ & -16.116 \\
 & $^{208}$Pb$(^{16}$O,$^{15}$C$)^{209}$Po & $-2p+1n$ & -19.993 \\
 & $^{208}$Pb$(^{16}$O,$^{16}$C$)^{208}$Po & $-2p+2n$ & -22.710 \\\hline\hline
\end{tabular}
\end{table}
Transfer probabilities for processes with different $\Delta Z$ are extracted by gating on the particular region of interest in the $\Delta E-E$ spectra, and normalizing the number of events to the total number of counts in the two forward angle monitor detectors. Overall normalization of the probabilities was achieved using the total quasi-elastic excitation function, following the procedure detailed in Ref.~\cite{Evers10}. Supplementary data comes from measurements using only a Si detector, and by integrating the number of counts in the total kinetic energy loss spectrum in a fixed energy interval. This is possible because of the well separated reaction $Q$-values for the predominant (as determined by our measurements and previous work) transfer processes in the reaction $^{16}$O+$^{208}$Pb (see Table~\ref{tab:qvalues}). Probabilities for the $\Delta Z=1$ ($1p$-stripping) and $\Delta Z=2$ transfer events are shown in Fig.~\ref{fig:transfer_rmin} by the filled squares and diamonds, respectively. The transfer probabilities are plotted as a function of the distance of closest approach assuming a Coulomb trajectory \cite{BrogliaWinther81}
\begin{equation}\label{eq:rc}
r_\text{min} = \frac{Z_p Z_t e^2}{4\pi\epsilon_0}\frac{1}{2E_{c.m.}}\left(1+\cosec{\frac{\theta_{c.m.}}{2}}\right)\,,
\end{equation}
where $Z_p,Z_t$ are the atomic number of projectile and target nucleus, and $E_{c.m.}$ and $\theta_{c.m.}$ are the energy and scattering angle in the centre-of-mass frame, respectively. The absolute probabilities at an energy around the fusion barrier agree very well with previous measurements \cite{Videbaek77} at $E_{c.m.}/V_B\sim 1.0$, which are shown by the large open square and diamond symbol for the N and C PLFs, respectively. Earlier measurements at the ANU of the N and C PLFs from Ref.~\cite{Timmers96} are shown by the smaller open squares and diamonds, respectively, and also show excellent agreement. Neither measurements however allowed a separation in mass of the PLFs, and it was commonly assumed that $\alpha$-particle transfer was the dominant $\Delta Z=2$ transfer process \cite{deVries75,Hasan79,Thompson89,Tamura80}.\par

In order to obtain insights into the $\Delta Z=2$ transfer mechanisms, an unambiguous identification of the dominant $\Delta Z=2$ transfer process is important. Reaction $Q$-values for $2p$ and $\alpha$-particle stripping are too similar to allow for a separation of these processes solely based on kinematic considerations, as is the case e.g. for the $1p$ and $1p1n$ stripping reactions (see Table~\ref{tab:qvalues}). However, $^{12}$C and $^{14}$C ions lose a different amount of energy in the gas of the detector telescope. The dashed curves in Fig.~\ref{fig:comparison} show the calculated energy losses for the C isotopes $^{12,13,14}$C. The locus of the majority of the measured $\Delta Z=2$ events coincides with the energy loss curve for $^{14}$C. This suggests that the majority of $\Delta Z=2$ events originate from the $2p$ transfer reaction leading to $^{14}$C, with a secondary contribution from $2p1n$ and $\alpha$-particle transfer. However, the unique identification of events with a particular transfer process depends critically on the accuracy of the energy loss calculations, which in turn depend on the modelling of the detector and the accuracy of the stopping power tables used.\par

\begin{figure}
\centering
\includegraphics[scale=0.45]{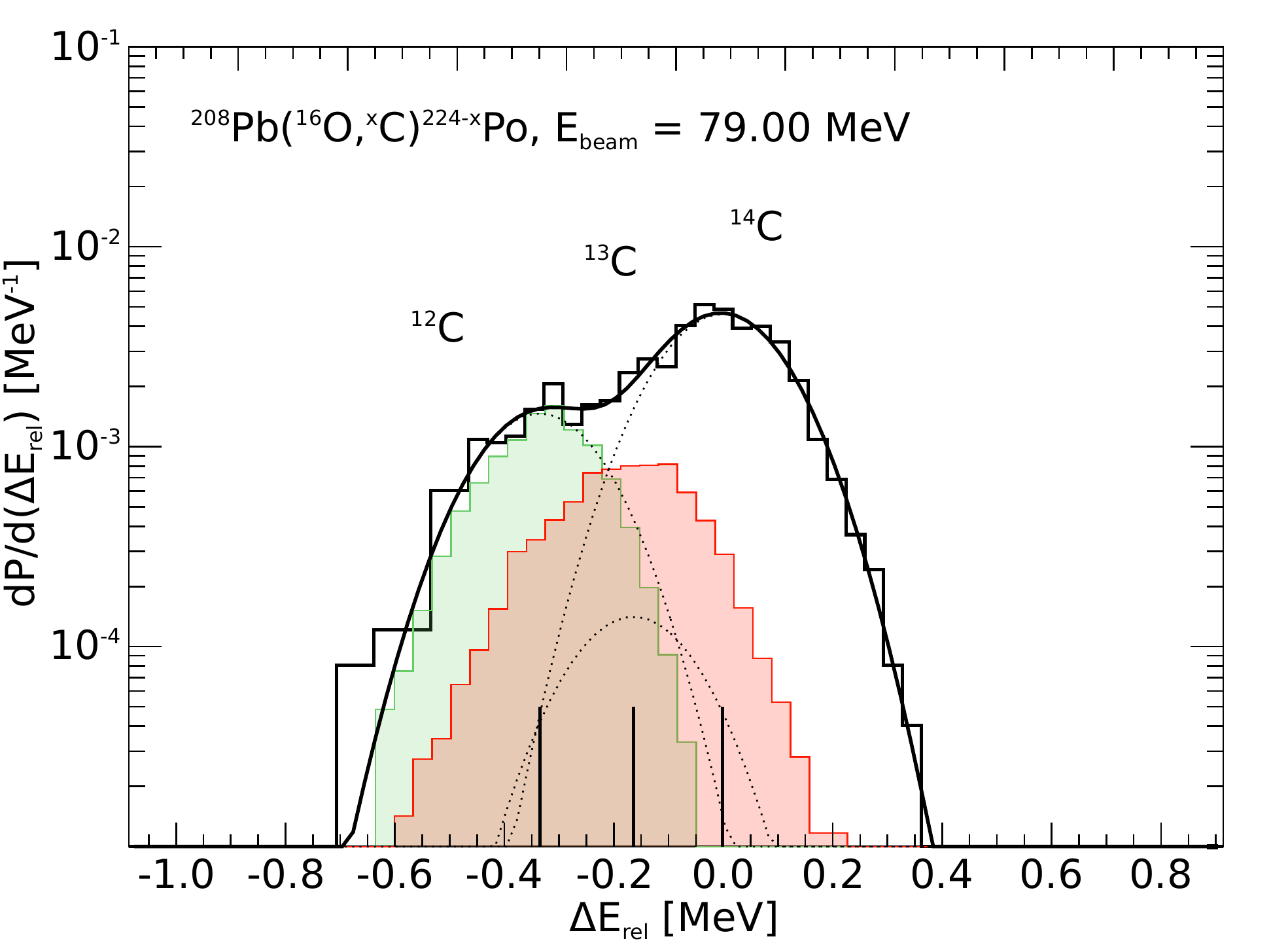}
\caption{(Color online) Relative energy loss $\Delta E_\text{rel}$ of the $\Delta Z=2$ transfer events (thick black histogram) relative to the calculated energy loss of $^{14}$C (see text) for an incident oxygen beam on a lead target at the indicated energy. Histograms of the relative energy losses for the elastic scattering measurements using beams of $^{12}$C and $^{13}$C are shown by the green and red shaded areas. The Gaussians fits (dotted curves) correspond to $2p$, $2p1n$ and $\alpha$-particle transfer leading to $^{14}$C, $^{13}$C and $^{12}$C ejectiles; the envelope of the fits is shown by the solid curve. Calculated relative energy losses for the three indicated transfer reactions are indicated by the vertical lines.\label{fig:DErel}}
\end{figure}
Beams of $^{12,13}$C were used in the same experiment to determine the accuracy of the energy loss calculations, by scattering them from a thick tantalum target to give empirical energy loss curves for the detector telescope. These measurements were reproduced satisfactorily by calculations using an improved version of the code STROP3, which uses stopping powers from Ref.~\cite{Ziegler77}. Based on the energy loss calculations, a new quantity, the \emph{relative energy loss} $\Delta E_\text{rel}$, was then defined. This quantity corresponds to the difference between the measured energy loss of the $\Delta Z=2$ PLFs and the calculated energy loss of $^{14}$C. The $\Delta E_\text{rel}$ projections are independent of differences in kinetic energy of the PLFs, thus they present a useful tool for (1) identifying the dominant transfer processes, and (2) determining their corresponding absolute probabilities integrated over all final states in the residual nuclei. Fig.~\ref{fig:DErel} shows the elastically scattered $^{12}$C and $^{13}$C beam particles, shaded green and red, respectively. The centroids coincide with the calculated energy losses for $^{12}$C and $^{13}$C (indicated by the vertical lines), therefore confirming the accuracy of the energy loss calculations as shown by the dashed curves in Fig.~\ref{fig:comparison}. The relative energy loss spectrum for the $\Delta Z=2$ transfer events measured in the $^{16}$O induced reaction is shown by the thick histogram in Fig.~\ref{fig:DErel} for an incident $^{16}$O beam energy corresponding to $E_{c.m.}/V_B = 0.98$.  The majority of these $\Delta Z=2$ events lie at $\Delta E_\text{rel}$ values \emph{above} the centroids of the $^{12}$C and $^{13}$C events. This identifies the majority of these events with $^{14}$C ejectiles, expected to be produced following the $2p$-stripping reaction $^{208}$Pb$(^{16}$O,$^{14}$C$)^{210}$Po.\par

The contributions to the total $\Delta Z=2$ transfer probability from the three transfer reactions $2p$, $2p1n$ and $\alpha$-particle transfer were extracted by fitting a 3-Gaussian distribution to the $\Delta E_\text{rel}$ spectrum. The width of each individual Gaussian is fixed to the value of the width of the Gaussian-shaped elastically scattered $^{12}$C distribution (green histogram in Fig.~\ref{fig:DErel}), and the relative energy losses between the three C isotopes are fixed to the values of the calculated energy losses for these particles. The envelope of the 3-Gaussian distribution as well as its individual Gaussian components are shown by the black solid and dotted curves in Fig.~\ref{fig:DErel}. Absolute probabilities for each $\Delta Z=2$ transfer process were then obtained by evaluating the integral of the individual Gaussian functions of the 3-Gaussian distribution and normalizing the sum of the three transfer probability components to the total $\Delta Z=2$ transfer probability.\par

Extracted probabilities for the two transfer processes $^{208}$Pb$(^{16}$O,$^{14}$C$)^{210}$Bi and $^{208}$Pb$(^{16}$O,$^{12}$C$)^{212}$Bi are shown in Fig.~\ref{fig:transfer_rmin} by the orange and green triangles, respectively. Transfer probabilities for the $2p1n$ stripping process $^{208}$Pb$(^{16}$O,$^{13}$C$)^{211}$Bi are not shown since they are at least 10 times smaller than those for $\alpha$-particle transfer (see Fig.~\ref{fig:DErel}). At sub-barrier energies, $2p$ transfer (orange triangles in Fig.~\ref{fig:transfer_rmin}) is the dominant process. $\alpha$-particle transfer probabilities (green triangles) are smaller by a factor of $\sim 2-3$ compared to that of $2p$ transfer. The difference in probabilities between $2p$ and $\alpha$ transfer increases with increasing beam energy, and is largest at $E_{c.m.}/V_{B}\sim 1.0$.\par

Insights into the transfer mechanisms and the significance of pairing correlations may be obtained by investigating the radial dependence of the average transfer form factor at large separation distances (i.e. for $r_\text{min}>r_B$, where $r_B$ is the fusion barrier radius). The asymptotic behaviour of the probability for transfer process $i$ is given by \cite{Oertzen01,Corradi09}
\begin{equation}\label{eq:transferprob}
P_i\propto\exp{\left(-2\kappa_i r_\text{min}\right)}\,,
\end{equation}
where $\kappa_i$ is the slope parameter for transfer process $i$. The asymptotic behaviour for $1p$ transfer $P_{1p}$ is shown by the dotted blue line in Fig.~\ref{fig:transfer_rmin}. In the absence of pairing correlations between the transferred nucleons, the probability for sequential two-nucleon transfer can be estimated by the product of the single nucleon transfer probabilities. The resulting transfer probabilities for sequential uncorrelated $2p$ transfer $\left(P_{1p}\right)^2$ are shown by the dotted orange line in Fig.~\ref{fig:transfer_rmin}. A significant enhancement of the observed $2p$ transfer probability is observed, by about one order of magnitude, which suggests a strong pairing correlation between the two transferred protons.\par

In conclusion, the $1p$, $2p$ and $\alpha$ transfer processes in the reaction $^{16}$O+$^{208}$Pb have been investigated at energies far below the fusion barrier. No detailed measurements previously existed at these deep sub-barrier energies. It is found that $2p$ transfer is the dominant $\Delta Z=2$ transfer process, extending to energies $E_{c.m.}/V_B\sim 0.9$. Corresponding absolute transfer probabilities reach a maximum of $\sim 10\%$ at a beam energy around the fusion barrier energy. The $2p$ transfer probability is strongly enhanced compared to predictions assuming sequential transfer of uncorrelated nucleons. This indicates a strong pairing correlation between the two transferred protons and suggests the existence of a supercurrent of Cooper-pair protons between the $^{16}$O and $^{208}$Pb nuclei. The enhancement of the $2p$ transfer probabilities is consistent with measurements for $^{16}$O-induced reactions on closed neutron-shell and open proton-shell targets at energies near and above the fusion barrier \cite{Oertzen73}. The significance of $2p$ transfer already at energies \emph{well below} the fusion barrier demonstrates that pairing correlations in $^{16}$O may play a more important role than generally assumed.\par 

This will have significant implications for both model calculations of nuclear collisions as well as nuclear structure. Regarding the former, it has been a challenge for many decades to simultaneously reproduce all observables related to individual reaction processes (elastic scattering, transfer, fusion, fission) for the $^{16}$O+$^{208}$Pb reaction at near-barrier energies. The current measurements show that cluster transfer occurs with significant probability even at sub-barrier energies, and must be correctly included in nuclear reaction model calculations \cite{Thompson89,Morton99,Dasgupta07}. This may then lead to a full understanding of this reaction and the structure of its constituting nuclei.\par

M.E. would like to thank W. von Oertzen for his insightful remarks and comments after reading this Letter. The authors acknowledge the financial support of an Australian Research Council Discovery Grant. 

\bibliography{refs}

\begin{thebibliography}{10}%
\makeatletter
\providecommand \@ifxundefined [1]{%
 \ifx #1\undefined \expandafter \@firstoftwo
 \else \expandafter \@secondoftwo
\fi
}%
\providecommand \@ifnum [1]{%
 \ifnum #1\expandafter \@firstoftwo
 \else \expandafter \@secondoftwo
\fi
}%
\providecommand \enquote [1]{``#1''}%
\providecommand \bibnamefont  [1]{#1}%
\providecommand \bibfnamefont [1]{#1}%
\providecommand \citenamefont [1]{#1}%
\providecommand\href[0]{\@sanitize\@href}%
\providecommand\@href[1]{\endgroup\@@startlink{#1}\endgroup\@@href}%
\providecommand\@@href[1]{#1\@@endlink}%
\providecommand \@sanitize [0]{\begingroup\catcode`\&12\catcode`\#12\relax}%
\@ifxundefined \pdfoutput {\@firstoftwo}{%
 \@ifnum{\z@=\pdfoutput}{\@firstoftwo}{\@secondoftwo}%
}{%
 \providecommand\@@startlink[1]{\leavevmode\special{html:<a href="#1">}}%
 \providecommand\@@endlink[0]{\special{html:</a>}}%
}{%
 \providecommand\@@startlink[1]{%
  \leavevmode
  \pdfstartlink
   attr{/Border[0 0 1 ]/H/I/C[0 1 1]}%
   user{/Subtype/Link/A<</Type/Action/S/URI/URI(#1)>>}%
  \relax
 }%
 \providecommand\@@endlink[0]{\pdfendlink}%
}%
\providecommand \url  [0]{\begingroup\@sanitize \@url }%
\providecommand \@url [1]{\endgroup\@href {#1}{\urlprefix}}%
\providecommand \urlprefix [0]{URL }%
\providecommand \Eprint[0]{\href }%
\@ifxundefined \urlstyle {%
  \providecommand \doi [1]{doi:\discretionary{}{}{}#1}%
}{%
  \providecommand \doi [0]{doi:\discretionary{}{}{}\begingroup
  \urlstyle{rm}\Url }%
}%
\providecommand \doibase [0]{http://dx.doi.org/}%
\providecommand \Doi[1]{\href{\doibase#1}}%
\providecommand \bibAnnote [3]{%
  \BibitemShut{#1}%
  \begin{quotation}\noindent
    \textsc{Key:}\ #2\\\textsc{Annotation:}\ #3%
  \end{quotation}%
}%
\providecommand \bibAnnoteFile [2]{%
  \IfFileExists{#2}{\bibAnnote {#1} {#2} {\input{#2}}}{}%
}%
\providecommand \typeout [0]{\immediate \write \m@ne }%
\providecommand \selectlanguage [0]{\@gobble}%
\providecommand \bibinfo [0]{\@secondoftwo}%
\providecommand \bibfield [0]{\@secondoftwo}%
\providecommand \translation [1]{[#1]}%
\providecommand \BibitemOpen[0]{}%
\providecommand \bibitemStop [0]{}%
\providecommand \bibitemNoStop [0]{.\EOS\space}%
\providecommand \EOS [0]{\spacefactor3000\relax}%
\providecommand \BibitemShut [1]{\csname bibitem#1\endcsname}%
\bibitem{Dasgupta98}%
  \BibitemOpen
  \bibfield{author}{%
  \bibinfo {author} {\bibfnamefont{M.}~\bibnamefont{Dasgupta}}, \bibinfo
  {author} {\bibfnamefont{D.~J.}\ \bibnamefont{Hinde}}, \bibinfo {author}
  {\bibfnamefont{N.}~\bibnamefont{Rowley}},\ and\ \bibinfo {author}
  {\bibfnamefont{A.~M.}\ \bibnamefont{Stefanini}},\ }%
  \bibfield{journal}{%
  \Doi{10.1146/annurev.nucl.48.1.401}{\bibinfo {journal} {Annu. Rev. Nucl.
  Part. Sci.}}\ }%
  \textbf{\bibinfo {volume} {48}},\ \bibinfo {pages} {401} (\bibinfo {year}
  {1998})%
  \bibAnnoteFile{NoStop}{Dasgupta98}%
\bibitem{Oertzen01}%
  \BibitemOpen
  \bibfield{author}{%
  \bibinfo {author} {\bibfnamefont{W.}~\bibnamefont{von Oertzen}}\ and\
  \bibinfo {author} {\bibfnamefont{A.}~\bibnamefont{Vitturi}},\ }%
  \bibfield{journal}{%
  \Doi{10.1088/0034-4885/64/10/202}{\bibinfo {journal} {Rep. Prog. Phys.}}\ }%
  \textbf{\bibinfo {volume} {64}},\ \bibinfo {pages} {1247} (\bibinfo {year}
  {2001})%
  \bibAnnoteFile{NoStop}{Oertzen01}%
\bibitem{Campbell00}%
  \BibitemOpen
  \bibfield{author}{%
  \bibinfo {author} {\bibfnamefont{E.~E.~B.}\ \bibnamefont{Campbell}}\ and\
  \bibinfo {author} {\bibfnamefont{F.}~\bibnamefont{Rohmund}},\ }%
  \bibfield{journal}{%
  \Doi{10.1088/0034-4885/63/7/202}{\bibinfo {journal} {Rep. Prog. Phys.}}\ }%
  \textbf{\bibinfo {volume} {63}},\ \bibinfo {pages} {1061} (\bibinfo {year}
  {2000})%
  \bibAnnoteFile{NoStop}{Campbell00}%
\bibitem{Broglia78}%
  \BibitemOpen
  \bibfield{author}{%
  \bibinfo {author} {\bibfnamefont{R.~A.}\ \bibnamefont{Broglia}}, \bibinfo
  {author} {\bibfnamefont{C.~H.}\ \bibnamefont{Dasso}}, \bibinfo {author}
  {\bibfnamefont{S.}~\bibnamefont{Landowne}}, \bibinfo {author}
  {\bibfnamefont{B.~S.}\ \bibnamefont{Nilsson}},\ and\ \bibinfo {author}
  {\bibfnamefont{A.}~\bibnamefont{Winther}},\ }%
  \bibfield{journal}{%
  \Doi{10.1016/0370-2693(78)90750-5}{\bibinfo {journal} {Phys. Lett. B}}\ }%
  \textbf{\bibinfo {volume} {73}},\ \bibinfo {pages} {401} (\bibinfo {year}
  {1978})%
  \bibAnnoteFile{NoStop}{Broglia78}%
\bibitem{Chiodi82}%
  \BibitemOpen
  \bibfield{author}{%
  \bibinfo {author} {\bibfnamefont{I.}~\bibnamefont{Chiodi}}, \bibinfo {author}
  {\bibfnamefont{S.}~\bibnamefont{Lunardi}}, \bibinfo {author}
  {\bibfnamefont{M.}~\bibnamefont{Morando}}, \bibinfo {author}
  {\bibfnamefont{C.}~\bibnamefont{Signorini}}, \bibinfo {author}
  {\bibfnamefont{G.}~\bibnamefont{Fortuna}}, \bibinfo {author}
  {\bibfnamefont{W.}~\bibnamefont{Starzecki}}, \bibinfo {author}
  {\bibfnamefont{A.}~\bibnamefont{Stefanini}}, \bibinfo {author}
  {\bibfnamefont{G.}~\bibnamefont{Korschinek}}, \bibinfo {author}
  {\bibfnamefont{H.}~\bibnamefont{Morinaga}}, \bibinfo {author}
  {\bibfnamefont{E.}~\bibnamefont{Nolte}},\ and\ \bibinfo {author}
  {\bibfnamefont{W.}~\bibnamefont{Schollmeier}},\ }%
  \bibfield{journal}{%
  \Doi{10.1007/BF02725528}{\bibinfo {journal} {Lett. Nuovo Cimento}}\ }%
  \textbf{\bibinfo {volume} {33}},\ \bibinfo {pages} {159} (\bibinfo {year}
  {1982})%
  \bibAnnoteFile{NoStop}{Chiodi82}%
\bibitem{Mermaz96}%
  \BibitemOpen
  \bibfield{author}{%
  \bibinfo {author} {\bibfnamefont{M.~C.}\ \bibnamefont{Mermaz}}\ and\ \bibinfo
  {author} {\bibfnamefont{M.}~\bibnamefont{Girod}},\ }%
  \bibfield{journal}{%
  \Doi{10.1103/PhysRevC.53.1819}{\bibinfo {journal} {Phys. Rev. C}}\ }%
  \textbf{\bibinfo {volume} {53}},\ \bibinfo {pages} {1819} (\bibinfo {year}
  {1996})%
  \bibAnnoteFile{NoStop}{Mermaz96}%
\bibitem{Corradi09}%
  \BibitemOpen
  \bibfield{author}{%
  \bibinfo {author} {\bibfnamefont{L.}~\bibnamefont{Corradi}}, \bibinfo
  {author} {\bibfnamefont{G.}~\bibnamefont{Pollarolo}},\ and\ \bibinfo {author}
  {\bibfnamefont{S.}~\bibnamefont{Szilner}},\ }%
  \bibfield{journal}{%
  \Doi{10.1088/0954-3899/36/11/113101}{\bibinfo {journal} {J. Phys. G: Nucl.
  Part. Phys.}}\ }%
  \textbf{\bibinfo {volume} {36}},\ \bibinfo {pages} {113101} (\bibinfo {year}
  {2009})%
  \bibAnnoteFile{NoStop}{Corradi09}%
\bibitem{Goldanskii65}%
  \BibitemOpen
  \bibfield{author}{%
  \bibinfo {author} {\bibfnamefont{V.~I.}\ \bibnamefont{Goldanskii}},\ }%
  \bibfield{journal}{%
  \Doi{10.1016/0031-9163(65)90603-7}{\bibinfo {journal} {Physics Letters}}\ }%
  \textbf{\bibinfo {volume} {14}},\ \bibinfo {pages} {233} (\bibinfo {year}
  {1965})%
  \bibAnnoteFile{NoStop}{Goldanskii65}%
\bibitem{Dietrich71}%
  \BibitemOpen
  \bibfield{author}{%
  \bibinfo {author} {\bibfnamefont{K.}~\bibnamefont{Dietrich}},\ }%
  \bibfield{journal}{%
  \Doi{10.1016/0003-4916(71)90067-4}{\bibinfo {journal} {Ann. Phys.}}\ }%
  \textbf{\bibinfo {volume} {66}},\ \bibinfo {pages} {480} (\bibinfo {year}
  {1971})%
  \bibAnnoteFile{NoStop}{Dietrich71}%
\bibitem{deVries75}%
  \BibitemOpen
  \bibfield{author}{%
  \bibinfo {author} {\bibfnamefont{R.~M.}\ \bibnamefont{DeVries}}, \bibinfo
  {author} {\bibfnamefont{D.}~\bibnamefont{Shapira}}, \bibinfo {author}
  {\bibfnamefont{W.~G.}\ \bibnamefont{Davies}}, \bibinfo {author}
  {\bibfnamefont{G.~C.}\ \bibnamefont{Ball}}, \bibinfo {author}
  {\bibfnamefont{J.~S.}\ \bibnamefont{Forster}},\ and\ \bibinfo {author}
  {\bibfnamefont{W.}~\bibnamefont{McLatchie}},\ }%
  \bibfield{journal}{%
  \Doi{10.1103/PhysRevLett.35.835}{\bibinfo {journal} {Phys. Rev. Lett.}}\ }%
  \textbf{\bibinfo {volume} {35}},\ \bibinfo {pages} {835} (\bibinfo {year}
  {1975})%
  \bibAnnoteFile{NoStop}{deVries75}%
\bibitem{Hasan79}%
  \BibitemOpen
  \bibfield{author}{%
  \bibinfo {author} {\bibfnamefont{H.}~\bibnamefont{Hasan}}\ and\ \bibinfo
  {author} {\bibfnamefont{C.~S.}\ \bibnamefont{Warke}},\ }%
  \bibfield{journal}{%
  \Doi{10.1016/0375-9474(79)90665-1}{\bibinfo {journal} {Nucl. Phys. A}}\ }%
  \textbf{\bibinfo {volume} {318}},\ \bibinfo {pages} {523} (\bibinfo {year}
  {1979})%
  \bibAnnoteFile{NoStop}{Hasan79}%
\bibitem{Thompson89}%
  \BibitemOpen
  \bibfield{author}{%
  \bibinfo {author} {\bibfnamefont{I.~J.}\ \bibnamefont{Thompson}}, \bibinfo
  {author} {\bibfnamefont{M.~A.}\ \bibnamefont{Nagarajan}}, \bibinfo {author}
  {\bibfnamefont{J.~S.}\ \bibnamefont{Lilley}},\ and\ \bibinfo {author}
  {\bibfnamefont{M.~J.}\ \bibnamefont{Smithson}},\ }%
  \bibfield{journal}{%
  \Doi{10.1016/0375-9474(89)90417-X}{\bibinfo {journal} {Nucl. Phys. A}}\ }%
  \textbf{\bibinfo {volume} {505}},\ \bibinfo {pages} {84} (\bibinfo {year}
  {1989})%
  \bibAnnoteFile{NoStop}{Thompson89}%
\bibitem{Tamura80}%
  \BibitemOpen
  \bibfield{author}{%
  \bibinfo {author} {\bibfnamefont{T.}~\bibnamefont{Tamura}}, \bibinfo {author}
  {\bibfnamefont{T.}~\bibnamefont{Udagawa}},\ and\ \bibinfo {author}
  {\bibfnamefont{M.~C.}\ \bibnamefont{Mermaz}},\ }%
  \bibfield{journal}{%
  \Doi{10.1016/0370-1573(80)90035-6}{\bibinfo {journal} {Phys. Rep.}}\ }%
  \textbf{\bibinfo {volume} {65}},\ \bibinfo {pages} {345} (\bibinfo {year}
  {1980})%
  \bibAnnoteFile{NoStop}{Tamura80}%
\bibitem{Becchetti74}%
  \BibitemOpen
  \bibfield{author}{%
  \bibinfo {author} {\bibfnamefont{F.~D.}\ \bibnamefont{Becchetti}}, \bibinfo
  {author} {\bibfnamefont{D.~G.}\ \bibnamefont{Kovar}}, \bibinfo {author}
  {\bibfnamefont{B.~G.}\ \bibnamefont{Harvey}}, \bibinfo {author}
  {\bibfnamefont{D.~L.}\ \bibnamefont{Hendrie}}, \bibinfo {author}
  {\bibfnamefont{H.}~\bibnamefont{Homeyer}}, \bibinfo {author}
  {\bibfnamefont{J.}~\bibnamefont{Mahoney}}, \bibinfo {author}
  {\bibfnamefont{W.}~\bibnamefont{von Oertzen}},\ and\ \bibinfo {author}
  {\bibfnamefont{N.~K.}\ \bibnamefont{Glendenning}},\ }%
  \bibfield{journal}{%
  \Doi{10.1103/PhysRevC.9.1543}{\bibinfo {journal} {Phys. Rev. C}}\ }%
  \textbf{\bibinfo {volume} {9}},\ \bibinfo {pages} {1543} (\bibinfo {year}
  {1974})%
  \bibAnnoteFile{NoStop}{Becchetti74}%
\bibitem{Goetz75}%
  \BibitemOpen
  \bibfield{author}{%
  \bibinfo {author} {\bibfnamefont{U.}~\bibnamefont{G\"{o}tz}}, \bibinfo
  {author} {\bibfnamefont{M.}~\bibnamefont{Ichimura}}, \bibinfo {author}
  {\bibfnamefont{R.~A.}\ \bibnamefont{Broglia}},\ and\ \bibinfo {author}
  {\bibfnamefont{A.}~\bibnamefont{Winther}},\ }%
  \bibfield{journal}{%
  \Doi{10.1016/0370-1573(75)90040-X}{\bibinfo {journal} {Phys. Rep.}}\ }%
  \textbf{\bibinfo {volume} {16}},\ \bibinfo {pages} {115} (\bibinfo {year}
  {1975})%
  \bibAnnoteFile{NoStop}{Goetz75}%
\bibitem{Videbaek77}%
  \BibitemOpen
  \bibfield{author}{%
  \bibinfo {author} {\bibfnamefont{F.}~\bibnamefont{Videbaek}}, \bibinfo
  {author} {\bibfnamefont{R.~B.}\ \bibnamefont{Goldstein}}, \bibinfo {author}
  {\bibfnamefont{L.}~\bibnamefont{Grodzins}}, \bibinfo {author}
  {\bibfnamefont{S.~G.}\ \bibnamefont{Steadman}}, \bibinfo {author}
  {\bibfnamefont{T.~A.}\ \bibnamefont{Belote}},\ and\ \bibinfo {author}
  {\bibfnamefont{J.~D.}\ \bibnamefont{Garrett}},\ }%
  \bibfield{journal}{%
  \Doi{10.1103/PhysRevC.15.954}{\bibinfo {journal} {Phys. Rev. C}}\ }%
  \textbf{\bibinfo {volume} {15}},\ \bibinfo {pages} {954} (\bibinfo {year}
  {1977})%
  \bibAnnoteFile{NoStop}{Videbaek77}%
\bibitem{Simenel10}%
  \BibitemOpen
  \bibfield{author}{%
  \bibinfo {author} {\bibfnamefont{C.}~\bibnamefont{Simenel}},\ }%
  \bibfield{journal}{%
  \Doi{10.1103/PhysRevLett.105.192701}{\bibinfo {journal} {Phys. Rev. Lett.}}\
  }%
  \textbf{\bibinfo {volume} {105}},\ \bibinfo {pages} {192701} (\bibinfo {year}
  {2010})%
  \bibAnnoteFile{NoStop}{Simenel10}%
\bibitem{Onderwater97}%
  \BibitemOpen
  \bibfield{author}{%
  \bibinfo {author} {\bibfnamefont{C.~J.~G.}\ \bibnamefont{Onderwater}},
  \bibinfo {author} {\bibfnamefont{K.}~\bibnamefont{Allaart}}, \bibinfo
  {author} {\bibfnamefont{E.~C.}\ \bibnamefont{Aschenauer}}, \bibinfo {author}
  {\bibfnamefont{T.~S.}\ \bibnamefont{Bauer}}, \bibinfo {author}
  {\bibfnamefont{D.~J.}\ \bibnamefont{Boersma}}, \bibinfo {author}
  {\bibfnamefont{E.}~\bibnamefont{Cisbani}}, \bibinfo {author}
  {\bibfnamefont{S.}~\bibnamefont{Frullani}}, \bibinfo {author}
  {\bibfnamefont{F.}~\bibnamefont{Garibaldi}}, \bibinfo {author}
  {\bibfnamefont{W.~J.~W.}\ \bibnamefont{Geurts}}, \bibinfo {author}
  {\bibfnamefont{D.~L.}\ \bibnamefont{Groep}}, \bibinfo {author}
  {\bibfnamefont{W.~H.~A.}\ \bibnamefont{Hesselink}}, \bibinfo {author}
  {\bibfnamefont{M.}~\bibnamefont{Iodice}}, \bibinfo {author}
  {\bibfnamefont{E.}~\bibnamefont{Jans}}, \bibinfo {author}
  {\bibfnamefont{N.}~\bibnamefont{Kalantar-Nayestanaki}}, \bibinfo {author}
  {\bibfnamefont{W.-J.}\ \bibnamefont{Kasdorp}}, \bibinfo {author}
  {\bibfnamefont{C.}~\bibnamefont{Kormanyos}}, \bibinfo {author}
  {\bibfnamefont{L.}~\bibnamefont{Lapik\'as}}, \bibinfo {author}
  {\bibfnamefont{J.~J.}\ \bibnamefont{van Leeuwe}}, \bibinfo {author}
  {\bibfnamefont{R.}~\bibnamefont{De~Leo}}, \bibinfo {author}
  {\bibfnamefont{A.}~\bibnamefont{Misiejuk}}, \bibinfo {author}
  {\bibfnamefont{A.~R.}\ \bibnamefont{Pellegrino}}, \bibinfo {author}
  {\bibfnamefont{R.}~\bibnamefont{Perrino}}, \bibinfo {author}
  {\bibfnamefont{R.}~\bibnamefont{Starink}}, \bibinfo {author}
  {\bibfnamefont{M.}~\bibnamefont{Steenbakkers}}, \bibinfo {author}
  {\bibfnamefont{G.}~\bibnamefont{van~der Steenhoven}}, \bibinfo {author}
  {\bibfnamefont{J.~J.~M.}\ \bibnamefont{Steijger}}, \bibinfo {author}
  {\bibfnamefont{M.~A.}\ \bibnamefont{van Uden}}, \bibinfo {author}
  {\bibfnamefont{G.~M.}\ \bibnamefont{Urciuoli}}, \bibinfo {author}
  {\bibfnamefont{L.~B.}\ \bibnamefont{Weinstein}},\ and\ \bibinfo {author}
  {\bibfnamefont{H.~W.}\ \bibnamefont{Willering}},\ }%
  \bibfield{journal}{%
  \Doi{10.1103/PhysRevLett.78.4893}{\bibinfo {journal} {Phys. Rev. Lett.}}\ }%
  \textbf{\bibinfo {volume} {78}},\ \bibinfo {pages} {4893} (\bibinfo {year}
  {1997})%
  \bibAnnoteFile{NoStop}{Onderwater97}%
\bibitem{Hesselink00}%
  \BibitemOpen
  \bibfield{author}{%
  \bibinfo {author} {\bibfnamefont{W.~H.~A.}\ \bibnamefont{Hesselink}},
  \bibinfo {author} {\bibfnamefont{R.}~\bibnamefont{Starink}}, \bibinfo
  {author} {\bibfnamefont{W.~H.~A.}\ \bibnamefont{Hesselink}}, \bibinfo
  {author} {\bibfnamefont{D.~L.}\ \bibnamefont{Groep}}, \bibinfo {author}
  {\bibfnamefont{E.}~\bibnamefont{Jans}},\ and\ \bibinfo {author}
  {\bibfnamefont{R.}~\bibnamefont{Starink}},\ }%
  \bibfield{journal}{%
  \Doi{10.1016/S0146-6410(00)00062-4}{\bibinfo {journal} {Prog. Part. Nucl.
  Phys.}}\ }%
  \textbf{\bibinfo {volume} {44}},\ \bibinfo {pages} {89} (\bibinfo {year}
  {2000})%
  \bibAnnoteFile{NoStop}{Hesselink00}%
\bibitem{Dickhoff10}%
  \BibitemOpen
  \bibfield{author}{%
  \bibinfo {author} {\bibfnamefont{W.~H.}\ \bibnamefont{Dickhoff}},\ }%
  \bibfield{journal}{%
  \Doi{10.1088/0954-3899/37/6/064007}{\bibinfo {journal} {J. Phys. G: Nucl.
  Part. Phys.}}\ }%
  \textbf{\bibinfo {volume} {37}},\ \bibinfo {pages} {1} (\bibinfo {year}
  {2010})%
  \bibAnnoteFile{NoStop}{Dickhoff10}%
\bibitem{Assie09}%
  \BibitemOpen
  \bibfield{author}{%
  \bibinfo {author} {\bibfnamefont{M.}~\bibnamefont{Assi\'{e}}}\ and\ \bibinfo
  {author} {\bibfnamefont{D.}~\bibnamefont{Lacroix}},\ }%
  \bibfield{journal}{%
  \Doi{10.1103/PhysRevLett.102.202501}{\bibinfo {journal} {Phys. Rev. Lett.}}\
  }%
  \textbf{\bibinfo {volume} {102}},\ \bibinfo {pages} {202501} (\bibinfo {year}
  {2009})%
  \bibAnnoteFile{NoStop}{Assie09}%
\bibitem{Abdullah03}%
  \BibitemOpen
  \bibfield{author}{%
  \bibinfo {author} {\bibfnamefont{M.~N.~A.}\ \bibnamefont{Abdullah}}, \bibinfo
  {author} {\bibfnamefont{S.}~\bibnamefont{Hossain}}, \bibinfo {author}
  {\bibfnamefont{M.~S.~I.}\ \bibnamefont{Sarker}}, \bibinfo {author}
  {\bibfnamefont{S.~K.}\ \bibnamefont{Das}}, \bibinfo {author}
  {\bibfnamefont{A.~S.~B.}\ \bibnamefont{Tariq}}, \bibinfo {author}
  {\bibfnamefont{M.~A.}\ \bibnamefont{Uddin}}, \bibinfo {author}
  {\bibfnamefont{A.~K.}\ \bibnamefont{Basak}}, \bibinfo {author}
  {\bibfnamefont{S.}~\bibnamefont{Ali}}, \bibinfo {author}
  {\bibfnamefont{H.~M.~S.}\ \bibnamefont{Gupta}},\ and\ \bibinfo {author}
  {\bibfnamefont{F.~B.}\ \bibnamefont{Malik}},\ }%
  \bibfield{journal}{%
  \Doi{10.1140/epja/i2002-10168-7}{\bibinfo {journal} {Eur. Phys. J. A}}\ }%
  \textbf{\bibinfo {volume} {18}},\ \bibinfo {pages} {65} (\bibinfo {year}
  {2003})%
  \bibAnnoteFile{NoStop}{Abdullah03}%
\bibitem{Dorsch08}%
  \BibitemOpen
  \bibfield{author}{%
  \bibinfo {author} {\bibfnamefont{T.}~\bibnamefont{Dorsch}}, \bibinfo {author}
  {\bibfnamefont{H.~G.}\ \bibnamefont{Bohlen}}, \bibinfo {author}
  {\bibfnamefont{W.}~\bibnamefont{von Oertzen}}, \bibinfo {author}
  {\bibfnamefont{R.}~\bibnamefont{Kr\"{u}cken}}, \bibinfo {author}
  {\bibfnamefont{T.}~\bibnamefont{Faestermann}}, \bibinfo {author}
  {\bibfnamefont{M.}~\bibnamefont{Mahgoub}}, \bibinfo {author}
  {\bibfnamefont{T.}~\bibnamefont{Kokalova}}, \bibinfo {author}
  {\bibfnamefont{C.}~\bibnamefont{Wheldon}}, \bibinfo {author}
  {\bibfnamefont{M.}~\bibnamefont{Milin}}, \bibinfo {author}
  {\bibfnamefont{H.}~\bibnamefont{Wirth}},\ and\ \bibinfo {author}
  {\bibfnamefont{R.}~\bibnamefont{Hertenberger}},\ }%
  \bibfield{journal}{%
  \Doi{10.1088/1742-6596/111/1/012039}{\bibinfo {journal} {Journal of Physics:
  Conference Series}}\ }%
  \textbf{\bibinfo {volume} {111}},\ \bibinfo {pages} {012039} (\bibinfo {year}
  {2008})%
  \bibAnnoteFile{NoStop}{Dorsch08}%
\bibitem{Timmers96}%
  \BibitemOpen
  \bibfield{author}{%
  \bibinfo {author} {\bibfnamefont{H.}~\bibnamefont{Timmers}},\ }%
  \emph{\bibinfo {title} {{Expressions of inner freedom}}},\ Ph.D. thesis,\
  \bibinfo {school} {The Australian National University} (\bibinfo {month}
  {06}\ \bibinfo {year} {1996}),\ \bibinfo {note}
  {\url{http://www.pems.adfa.edu.au/~s9471553/level1/pdffiles/thesisabstract.h%
tml}}%
  \bibAnnoteFile{NoStop}{Timmers96}%
\bibitem{Pieper78}%
  \BibitemOpen
  \bibfield{author}{%
  \bibinfo {author} {\bibfnamefont{S.~C.}\ \bibnamefont{Pieper}}, \bibinfo
  {author} {\bibfnamefont{M.~H.}\ \bibnamefont{Macfarlane}}, \bibinfo {author}
  {\bibfnamefont{D.~H.}\ \bibnamefont{Gloeckner}}, \bibinfo {author}
  {\bibfnamefont{D.~G.}\ \bibnamefont{Kovar}}, \bibinfo {author}
  {\bibfnamefont{D.}~\bibnamefont{Becchetti}}, \bibinfo {author}
  {\bibfnamefont{B.~G.}\ \bibnamefont{Harvey}}, \bibinfo {author}
  {\bibfnamefont{D.~L.}\ \bibnamefont{Hendrie}}, \bibinfo {author}
  {\bibfnamefont{H.}~\bibnamefont{Homeyer}}, \bibinfo {author}
  {\bibfnamefont{J.}~\bibnamefont{Mahoney}}, \bibinfo {author}
  {\bibfnamefont{F.}~\bibnamefont{Puehlhofer}}, \bibinfo {author}
  {\bibfnamefont{W.}~\bibnamefont{von Oertzen}},\ and\ \bibinfo {author}
  {\bibfnamefont{M.~S.}\ \bibnamefont{Zisman}},\ }%
  \bibfield{journal}{%
  \Doi{10.1103/PhysRevC.18.180}{\bibinfo {journal} {Phys. Rev. C}}\ }%
  \textbf{\bibinfo {volume} {18}},\ \bibinfo {pages} {180} (\bibinfo {year}
  {1978})%
  \bibAnnoteFile{NoStop}{Pieper78}%
\bibitem{Evers10}%
  \BibitemOpen
  \bibfield{author}{%
  \bibinfo {author} {\bibfnamefont{M.}~\bibnamefont{Evers}}, \bibinfo {author}
  {\bibfnamefont{D.~J.}\ \bibnamefont{Hinde}}, \bibinfo {author}
  {\bibfnamefont{M.}~\bibnamefont{Dasgupta}}, \bibinfo {author}
  {\bibfnamefont{D.~H.}\ \bibnamefont{Luong}}, \bibinfo {author}
  {\bibfnamefont{R.}~\bibnamefont{Rafiei}},\ and\ \bibinfo {author}
  {\bibfnamefont{R.}~\bibnamefont{du~Rietz}},\ }%
  \bibfield{journal}{%
  \Doi{10.1103/PhysRevC.81.014602}{\bibinfo {journal} {Phys. Rev. C}}\ }%
  \textbf{\bibinfo {volume} {81}},\ \bibinfo {pages} {014602} (\bibinfo {year}
  {2010})%
  \bibAnnoteFile{NoStop}{Evers10}%
\bibitem{BrogliaWinther81}%
  \BibitemOpen
  \bibfield{author}{%
  \bibinfo {author} {\bibfnamefont{R.~A.}\ \bibnamefont{Broglia}}\ and\
  \bibinfo {author} {\bibfnamefont{A.}~\bibnamefont{Winther}},\ }%
  \emph{\bibinfo {title} {{Heavy Ion Reactions (Lecture Notes), Volume 1:
  Elastic and Inelastic Reactions}}},\ Vol.~\bibinfo {volume} {1}\ (\bibinfo
  {publisher} {The Benjamin/Cummings Publishing Company, Inc.},\ \bibinfo
  {year} {1981})%
  \bibAnnoteFile{NoStop}{BrogliaWinther81}%
\bibitem{Ziegler77}%
  \BibitemOpen
  \bibfield{author}{%
  \bibinfo {author} {\bibfnamefont{J.~F.}\ \bibnamefont{Ziegler}},\ }%
  \emph{\bibinfo {title} {{Helium: Stopping Powers and Ranges in All Elemental
  Matter}}},\ \bibinfo {series} {The Stopping and Ranges of Ions in Matter},
  Vol.~\bibinfo {volume} {4}\ (\bibinfo {publisher} {Elsevier},\ \bibinfo
  {year} {1977})%
  \bibAnnoteFile{NoStop}{Ziegler77}%
\bibitem{Oertzen73}%
  \BibitemOpen
  \bibfield{author}{%
  \bibinfo {author} {\bibfnamefont{W.}~\bibnamefont{von Oertzen}}, \bibinfo
  {author} {\bibfnamefont{H.~G.}\ \bibnamefont{Bohlen}},\ and\ \bibinfo
  {author} {\bibfnamefont{B.}~\bibnamefont{Gebauer}},\ }%
  \bibfield{journal}{%
  \Doi{10.1016/0375-9474(73)90025-0}{\bibinfo {journal} {Nucl. Phys. A}}\ }%
  \textbf{\bibinfo {volume} {207}},\ \bibinfo {pages} {91} (\bibinfo {year}
  {1973})%
  \bibAnnoteFile{NoStop}{Oertzen73}%
\bibitem{Morton99}%
  \BibitemOpen
  \bibfield{author}{%
  \bibinfo {author} {\bibfnamefont{C.~R.}\ \bibnamefont{Morton}}, \bibinfo
  {author} {\bibfnamefont{A.~C.}\ \bibnamefont{Berriman}}, \bibinfo {author}
  {\bibfnamefont{M.}~\bibnamefont{Dasgupta}}, \bibinfo {author}
  {\bibfnamefont{D.~J.}\ \bibnamefont{Hinde}}, \bibinfo {author}
  {\bibfnamefont{J.~O.}\ \bibnamefont{Newton}}, \bibinfo {author}
  {\bibfnamefont{K.}~\bibnamefont{Hagino}},\ and\ \bibinfo {author}
  {\bibfnamefont{I.~J.}\ \bibnamefont{Thompson}},\ }%
  \bibfield{journal}{%
  \Doi{10.1103/PhysRevC.60.044608}{\bibinfo {journal} {Phys. Rev. C}}\ }%
  \textbf{\bibinfo {volume} {60}},\ \bibinfo {pages} {044608} (\bibinfo {year}
  {1999})%
  \bibAnnoteFile{NoStop}{Morton99}%
\bibitem{Dasgupta07}%
  \BibitemOpen
  \bibfield{author}{%
  \bibinfo {author} {\bibfnamefont{M.}~\bibnamefont{Dasgupta}}, \bibinfo
  {author} {\bibfnamefont{D.~J.}\ \bibnamefont{Hinde}}, \bibinfo {author}
  {\bibfnamefont{A.}~\bibnamefont{Diaz-Torres}}, \bibinfo {author}
  {\bibfnamefont{B.}~\bibnamefont{Bouriquet}}, \bibinfo {author}
  {\bibfnamefont{C.~I.}\ \bibnamefont{Low}}, \bibinfo {author}
  {\bibfnamefont{G.~J.}\ \bibnamefont{Milburn}},\ and\ \bibinfo {author}
  {\bibfnamefont{J.~O.}\ \bibnamefont{Newton}},\ }%
  \bibfield{journal}{%
  \Doi{10.1103/PhysRevLett.99.192701}{\bibinfo {journal} {Phys. Rev. Lett.}}\
  }%
  \textbf{\bibinfo {volume} {99}},\ \bibinfo {pages} {192701} (\bibinfo {year}
  {2007})%
  \bibAnnoteFile{NoStop}{Dasgupta07}%
\end{thebibliography}%

\end{document}